\documentclass[a4paper, 12pt]{article}

\usepackage{amsmath,amssymb,amsthm}

\usepackage[all]{xy}
\usepackage{latexsym}

\usepackage{enumitem}
\usepackage{algorithmic}
\usepackage{algorithm}

\makeatletter
\newcommand{\todo}[1]{\marginpar{\textbf{TODO\footnotemark}}\@latex@warning{TODO: #1}\footnotetext{ #1}}
\makeatother

\theoremstyle{plain}
\newtheorem{theorem}{Theorem} 
\newtheorem{lemma}[theorem]{Lemma} 
\newtheorem{definition}[theorem]{Definition} 
\newtheorem{example}[theorem]{Example} 
\newtheorem{corollary}[theorem]{Corollary}

\newcommand{\pow}{\mathcal{P}}

\newcommand{\Lo}{\mathsf{L}}
\newcommand{\ML}{\mathsf{M}}
\newcommand{\FL}{\mathsf{Fml}_\mathsf{L}}
\newcommand{\FM}{\mathsf{Fml}_\mathsf{M}}
\newcommand{\AK}{\mathsf{AK}}
\newcommand{\Sec}{\mathsf{Sec}}
\newcommand{\KB}{\mathsf{KB}}
\newcommand{\A}{\mathbb{A}}
\newcommand{\Cens}{\mathsf{Cens}}
\newcommand{\PK}{(\KB,\AK,\Sec)}
\newcommand{\CensPK}{\Cens_{\PK}}
\newcommand{\eval}{\mathsf{eval}}
\newcommand{\co}{\mathsf{cont}}
\newcommand{\full}{\mathsf{full}}
\newcommand{\M}{\mathcal{M}}

\title{ No-Go Theorems for Data Privacy}
\author{Thomas Studer\thanks{supported by the Swiss National Science Foundation grant 200020\_184625.}}
\date{Institute of Computer Science\\University of Bern\\Bern\\Switzerland\\ \tt{thomas.studer@inf.unibe.ch}}

\begin{document}
\maketitle


\begin{abstract}
Controlled query evaluation (CQE) is an approach to guarantee data privacy for database and knowledge base systems. CQE-systems feature a censor function that may distort the answer to a query  in order to hide sensitive information. We introduce a high-level formalization of controlled query evaluation and define several desirable properties of CQE-systems. Finally we establish two no-go theorems, which show that certain combinations of these properties cannot be obtained.

\end{abstract} 
\section{Introduction}

Controlled query evaluation (CQE) refers to a data privacy mechanism where 
the database (or knowledge base) is equipped with a censor function.
This censor checks for each query whether the answer to the query would reveal sensitive information to a user. If this is the case, then the censor will distort  the answer. 
Essentially, there are two possibilities how an answer may be distorted: 
\begin{enumerate}
\item
the CQE-system may refuse to answer the query~\cite{Sicherman:1983} or 
\item
the CQE-system may give an incorrect answer, i.e.~it lies~\cite{bokrsu95}.
\end{enumerate}
This censor based approach has the advantage that the task of maintaining privacy is separated from the task of keeping the data. This gives more flexibility than an integrated approach (like hiding rows in a database) and guarantees than no information is leaked through otherwise unidentified inference channels.
Controlled query evaluation has been applied to a variety of data models and control mechansims, see, e.g.~\cite{Biskup2000,BISKUP2001,BiskupB04,Biskup2004,Biskup2008,sw14}.

No-go theorems are well-known in theoretical physics where they describe particular situations that are not physically possible.  Often the term is used for results in quantum mechanics like Bell's theorem~\cite{PhysicsPhysiqueFizika.1.195}, the Kochen--Specker theorem~\cite{Koc}, or, for a more recent example,  the Frauchiger--Renner paradox~\cite{FrauchigerRenner}.
Nurgalieva and del Rio~
\cite{DBLP:journals/corr/abs-1804-01106} provide a modal logic analysis of the latter paradox.
Arrow's theorem~\cite{doi:10.1086/256963} in social choice theory also is a no-go theorem  stating that no voting system can be designed that meets certain given fairness conditions. 
Pacuit and Yang~
\cite{Pacuit2016}  present a version of independence logic in which Arrow's theorem is derivable.


In the present paper we develop a highly abstract model for dynamic query evaluation systems like CQE. We formulate several desirable properties of CQE-systems in our framework and establish two no-go theorems saying that certain combinations of those properties are impossible.
The main contribution of this paper is the presentation of the abstract logical framework as well as the high-level formulation of the no-go theorems. Note that some particular instances of our results have already been known~\cite{Biskup2000,sw14}.

There are many different notions of privacy available in the literature. For our results, we rely on  \emph{provable privacy}~\cite{StoffelStuder,StouppaStuder}, which is a rather weak notion of data privacy.
Note that using a weak definition of privacy makes our impossibility theorems actually stronger since they state that under certain conditions not even this weak form of privacy can be achieved.

Clearly our work is also connected to the issues of lying and deception. Logics dealing with these notions are introduced and studied, e.g., in~\cite{Ditmarsch18,Ditmarsch14,IcardPhD}.








%
%
%
%
%
%
%
%
%
%
%
%


\section{Logical Preliminaries}
Let $X$ be a set. We use $\pow(X)$ to denote the power set of $X$.
For sets $\Gamma$ and~$\Delta$ we use $\Gamma, \Delta$ for  $\Gamma \cup \Delta$. Moreover, in such a context we write $A$ for the singleton set $\{A\}$. Hence $\Gamma, A$ stands for $\Gamma \cup \{A\}$.

\begin{definition}
A \emph{logic $\Lo$} is given by
\begin{enumerate}
\item a set of formulas $\FL$ and 
\item a consequence relation $\vdash_\Lo$ for $\Lo$ that is a relation between sets of formulas and formulas, i.e.~$\vdash_\Lo \ \subseteq\ \pow(\FL) \times \FL$ satisfying
for all $A,C \in \FL$ and $\Gamma, \Delta \in \pow(\FL)$:
\begin{enumerate}
\item reflexivity: $\{A\} \vdash_\Lo A$;
\item weakening:  $\Gamma \vdash_\Lo A \ \Longrightarrow\ \Gamma,\Delta \vdash_\Lo A$;
\item transitivity: $\Gamma \vdash_\Lo C \text{ and } \Delta,C \vdash_\Lo A  \ \Longrightarrow\  \Gamma,\Delta \vdash_\Lo A$.
\end{enumerate}
\end{enumerate}
\end{definition}

Transitivity is sometimes called \emph{cut}.
The previous definition gives us  single conclusion consequence relations, which is sufficient for 
the purpose of this paper.  For other notions of consequence relations see, e.g.,~\cite{Avron:1991} and~\cite{Iemhoff2016}.

As usual, we write $\vdash_\Lo A$ for $ \emptyset \vdash_\Lo A$.
A formula $A$ is called a \emph{theorem of\/ $\Lo$} if $\vdash_\Lo A$.

We do not specify the logic $\Lo$ any further.  The only thing we need is a consequence relation as given above.
For instance, $\Lo$ may be classical propositional logic with $\vdash_\Lo$ being the usual derivation relation (see Section~\ref{sec:nonref:1}) or $\Lo$  may be a description logic with $\vdash_\Lo$ being its semantic consequence relation~\cite{sw14}.

\begin{definition}\ 
\begin{enumerate}
\item
A logic $\Lo$ is called \emph{consistent} if there exists a formula $A \in \FL$ such that
$\not{\vdash_\Lo} A$.
\item
A set $\Gamma$ of\/ $\FL$-formulas is called \emph{$\Lo$-consistent} if there exists a formula $A \in \FL$ such that
$\Gamma \not{\vdash_\Lo} A$.
\end{enumerate}
\end{definition}

We need a simple  modal logic $\ML$ over $\Lo$.

\begin{definition}
The set of formulas $\FM$ is given inductively by:
\begin{enumerate}
\item if $A$ is a formula of $\FL$, then $\Box A$  is a formula of $\FM$;
\item $\bot$ is a formula of $\FM$;
\item if $A$ and $B$ are  formulas of $\FM$, so is $A \to B$, too.
\end{enumerate}
\end{definition}

We define the remaining classical connectives $\top$, $\land$, $\lor$, and $\lnot$ as usual. Note that $\ML$ is 
not a fully-fledged modal logic. For instance, it does not include nested modalities.

We give semantics to $\FM$-formulas as follows.
\begin{definition}
An $\ML$-model $\M$ is a set of sets of\/ $\FL$-formulas, that is
\[
\M \subseteq \pow(\FL).
\]
\end{definition}

\begin{definition}
Let $\M$ be an $\ML$-model. Truth of an $\FM$-formula in~$\M$ is inductively defined by:
\begin{enumerate}
\item $\M \Vdash \Box A$ if{f} $w \vdash_\Lo A$ for all $w \in \M$;
\item $\M \not\Vdash \bot$;
\item $\M \Vdash A \to B$ if{f}  $\M \not\Vdash A$  or $\M \Vdash B$.
\end{enumerate}
\end{definition}

We use the following standard definition.
\begin{definition}
Let $\Gamma$ be a set of\/ $\FM$-formulas.
\begin{enumerate}
\item
We write $\M \Vdash \Gamma$ if{f} $\M \Vdash A$ for each $A \in \Gamma$.
\item
$\Gamma$ is called \emph{satisfiable} if{f}  there exists an $\ML$-model $\M$ with   $\M \Vdash \Gamma$.
\item
$\Gamma$ \emph{entails} a formula $A$, in symbols $\Gamma \models A$, if{f} for each model $\M$ we have that
\[
\M \Vdash \Gamma \quad\text{implies}\quad \M \Vdash A.
\]
\end{enumerate}
\end{definition}

%
%
%
%
%
%

\section{Privacy}

\begin{definition}\label{d:pk:1}
A \emph{privacy configuration} is a triple $\PK$ that consists of:
\begin{enumerate}
\item the knowledge base $\KB \subseteq \FL$, which is only accessible via the censor;
\item the set of a priori knowledge $\AK \subseteq \FM$, which formalizes general background knowledge known to the attacker and the censor;
\item the set of secrets $\Sec \subseteq \FL$, which should be protected by the censor.
\end{enumerate}
A privacy configuration $\PK$ satisfies the following conditions:
\begin{enumerate}
\item $\KB$ is $\Lo$-consistent (consistency);
\item $\{\KB\}  \Vdash \AK$ (truthful start);
\item $\AK \not\models \Box s$ for each $s \in \Sec$ (hidden secrets).
\end{enumerate}
\end{definition}

Note that in the above definition, $\KB$ and $\Sec$ are sets of $\FL$-formulas while $\AK$ is a set of $\FM$-formulas. Thus $\AK$ may not only contain domain knowledge but also knowledge about the structure of $\KB$. 
This is further explained in Section~\ref{sec:nonref:1}.

A \emph{query} to a knowledge base $\KB$ is simply a formula of $\FL$.  

Given a logic $\Lo$, 
we can evaluate a query $q$ over a knowledge base~$\KB$. 
There are two possible answers:  $t$ (true) and $u$ (unknown).
\begin{definition}
The evaluation function $\eval$ is defined by:
\[
\eval(\KB,q) := \begin{cases}
				t &\text{if}\quad  \KB \vdash_\Lo q \\
				u &\text{otherwise}
			\end{cases}		
\]
\end{definition}
If the language of the logic $\Lo$ includes negation, then one may also consider an evaluation function that can return the value $f$ (false), i.e.~one defines $\eval(\KB,q) := f$ if $\KB \vdash_\Lo \lnot q$. However, in the general setting of this paper, we cannot include this case.

A censor has to hide the secrets. In order to achieve this, it can not only answer $t$ and $u$ to a query but also $r$ (refuse to answer).
We denote the set of possible answers of a censor by 
\[
\A:=\{t,u,r\}.
\]

Let $X$ be a set. 
Then $X^\omega$ denotes the set of infinite sequences of elements of $X$.

\begin{definition}
A \emph{censor} is a mapping that assigns an answering function
\[
\CensPK : \FL^\omega \longrightarrow \A^\omega
\]
to each privacy configuration $\PK$.
By abuse of notation, we also call the answering function $\CensPK$ a \emph{censor}.
A sequence $q \in \FL^\omega$ is called \emph{query sequence}.
\end{definition}

Usually, the privacy configuration will be clear from the context. In that case we simply use  $\Cens$ instead of $\CensPK$.

Given a sequence $s$, we use $s_i$ to denote its $i$-th element. That is for a query sequence 
$q \in \FL^\omega$, we use $q_i$ to denote the $i$-th  query and $\Cens(q)_i$ to denote the $i$-th answer of the censor.

\begin{example}
Let $A,B,C \in \FL$. We define a privacy configuration with $\KB = \{A,C\}$, $\AK = \emptyset$, and $\Sec = \{C\}$.
 A censor $\Cens$ yields an answering function $\CensPK$, which applied to  a query sequence $q = (A,B,C,\ldots)$ yields a sequence of answers, e.g.,
\[
\CensPK(q) = {t,u,r,\ldots}.
\]
In this case, $\CensPK$ gives true answers since $\eval(\KB,A) = t$ and $\eval(\KB,B) = u$ and it protects the secret be refusing to answer the query $C$. 

Another option for the answering function would be to answer the third query with $u$, i.e., it would lie (instead of refuse to answer) in order to protect the secret. 

A further option would be to always refuse the answer, i.e.
\[
\CensPK(q) = {r,r,r,\ldots}.
\]
This, of course, would be a trivial (and useless) answering function that would, however, preserve all secrets.
\end{example}

In this paper, we will consider continuous censors only, which are given as follows.
\begin{definition}
A censor $\Cens$  is  
\emph{continuous} if{f} for each privacy configuration $\PK$  and for all query sequences 
$q,q' \in  \FL^\omega$ and all $n \in \omega$ we have that
\[
q|_n = q'|_n \quad\Longrightarrow\quad \CensPK(q)|_n = \CensPK(q')|_n
\]
where for an infinite sequence $s= (s_1,s_2,\ldots)$, we use $s|_n$ to denote the initial segment of $s$ of length $n$, i.e.~$s|_n = (s_1, \ldots, s_n)$.
\end{definition}
Continuity means that the answer of a censor to a query does not depend on future queries, see also Lemma~\ref{l:lemmaOne:1}.

A censor is called truthful if it does not lie.
\begin{definition}
A censor $\Cens$  is called 
\emph{truthful} if{f} for each privacy configuration $\PK$, all query sequences $q=(q_1,q_2,\ldots)$, and all sequences 
\[
(a_1,a_2,\ldots) = \CensPK(q)
\]
 we have that for all $i\in \omega$
\[
a_i = \eval(\KB,q_i) \quad\text{or}\quad a_i = r.
\]
\end{definition}
Hence a truthful censor may refuse to answer a query in order to protect a secret but it will not give an incorrect answer.


In the modal logic $\ML$ over $\Lo$, we can express what knowledge one can gain from the answers of a censor to a query. This is called the content of the answer.
\begin{definition}
Given an answer $a \in \A$ to a query $q \in \FL$, we define its \emph{content} as follows:
\begin{align*}
\co(q,t) &:= \Box q\\
\co(q,u) &:= \lnot \Box q\\
\co(q,r) &:= \top
\end{align*}
Assume that we are given a privacy configuration $\PK$ and a censor $\Cens$.  We define the content of the answers of the censor to a query sequence $q \in \FL^\omega$ up to $n \in \omega$ by
\[
\co(\CensPK(q),n) := \bigcup_{1 \leq i \leq n}\!\{\co(q_i,a_i) \} \cup \AK
\]
where $a=\CensPK(q)$.
Note that here we have also included  the a priori knowledge.
\end{definition}

The following is a trivial observation showing the role of continuity.
\begin{lemma}\label{l:lemmaOne:1}
Let $\Cens$ be a continuous censor.  The content function is monotone in the second argument:  for $m \leq n$ we have
\[
\co(\Cens(q),m) \subseteq \co(\Cens(q),n).
\]
\end{lemma}

We call a censor credible if it does not return contradicting answers.
\begin{definition}\label{def:credible:1}
A censor $\Cens$  is called 
\emph{credible}
if{f} for each privacy configuration $\PK$ and for every query sequence $q$ and every $n \in \omega$, 
the set 
$
\co(\CensPK(q),n) 
$
is satisfiable.
\end{definition}


\begin{definition}\label{d:fullcontent:1}
The \emph{full content} of a knowledge base $\KB$ is given by
\[
\full(\KB):= \bigcup_{A \in \FL} \co(A, \eval(\KB, A)).
\]
\end{definition}

\begin{lemma}
For any knowledge base $\KB$, we have that
\[
\{\KB\} \Vdash \full(\KB).
\]
\end{lemma}
\begin{proof}
Let $A$ be an $\FL$-formula. We dinstinguish:
\begin{enumerate}
\item $\KB \vdash_\Lo A$. Then $\Box A \in  \full(\KB)$ and  further $\{\KB\} \Vdash \Box A$. 
\item $\KB \not\vdash_\Lo A$. Then $\lnot \Box A \in  \full(\KB)$ and further $\{\KB\} \Vdash \lnot\Box A$. 
\end{enumerate}
\end{proof}

\begin{lemma}
We let $\PK$ be a privacy configuration. Further we let 
$\CensPK$ be a truthful censor. For every query sequence $q$ and $n \in \omega$, we have that
\[
\co(\CensPK(q),n) \subseteq \full(\KB)  \cup \{\top\} \cup \AK.
\]
\end{lemma}
\begin{proof}
By induction on $n$. The base case $n=0$ is trivial since
\[
\co(\CensPK(q),0) =   \AK.
\]

Induction step. 
Since $\Cens$ is truthful, we have 
\[
a_{n+1} \in \{r, \eval(\KB,q_{n+1})\}.
\]
We distinguish:
\begin{enumerate}
\item $a_{n+1} = r$. Then 
$
\co(\Cens(q),n+1) = \co(\Cens(q),n) \cup\{\top\}
$
and the claim follows immediately from the induction hypothesis.
\item $a_{n+1} =\eval(\KB,q_{n+1})$. Then
\begin{equation*}
\begin{split}
\co(\Cens(q),n+1) =\ &\co(\Cens(q),n) \ \cup \\
&\co(q_{n+1}, \eval(\KB,q_{n+1})).
\end{split}
\end{equation*}
The claim follows from the induction hypothesis and 
\[
\co(q_{n+1}, \eval(\KB,q_{n+1})) \in \full (\KB),
\]
which holds by Definition~\ref{d:fullcontent:1}.
\end{enumerate}
\end{proof}

The following corollary is a generalization of Cor.~30 in~\cite{sw14}.

\begin{corollary}\label{cor:cred:1}
Every truthful censor is credible.
\end{corollary}
\begin{proof}
Let $\PK$ be a privacy configuration and 
$\Cens$ be a truthful censor for it.
By Definition~\ref{d:pk:1}, we have $\{\KB\}  \Vdash \AK$.
Thus by the two previous  lemmas, we find that for each $n \in \omega$, 
\begin{align*}
\{\KB\} \Vdash\  &\full(\KB) \cup \{\top\} \cup \AK \\ 
\intertext{and}
&\full(\KB) \cup \{\top\} \cup \AK \supseteq \co(\Cens(q),n),
\end{align*}
that means 
$
\co(\Cens(q),n) 
$
 is satisfiable for each $n\in \omega$ and thus
$\Cens$ is credible.
\end{proof}

There are several properties that a `good' censor should fulfil. 
We call a censor  effective if it protects all secrets.
\begin{definition}\label{def:effective:1}
A censor $\Cens$ is called \emph{effective} if{f} for each privacy configuration $\PK$ and for every query sequence $q \in \FL^\omega$ and every $n \in \omega$, we have
\[
\co(\CensPK(q),n) \not\models \Box s \quad\text{for each $s \in \Sec$}
\]
\end{definition}

A `good' censor should only distort an answer to a query when it is absolutely necessary, i.e.~when giving the correct answer would leak a secret. We call such a censor minimally invasive.
\begin{definition}
Let $\Cens$ be an effective and credible censor.
This censor is called \emph{minimally invasive} if{f} for each privacy configuration $\PK$  and for each query sequence $q \in \FL^\omega$, 
we have that whenever 
\[
\CensPK(q)_i \neq \eval(\KB, q_i),
\]
 replacing 
\begin{equation*}\label{eq:replace:1}
\CensPK(q)_i
\quad
\text{with}
\quad 
\eval(\KB, q_i)
\end{equation*}
would lead to a violation of effectiveness or credibility, that is~for any censor $\CensPK'$ such that 
\[
\CensPK'(q)|_{i-1} = \CensPK(q)|_{i-1} 
\]
and 
\[
\CensPK'(q)_i = \eval(\KB, q_i)
\]
we  have
that for some $n$
\[
\co(\CensPK'(q),n) \models \Box s \quad\text{for some $s \in \Sec$}
\]
or 
\[
\co(\CensPK'(q),n) \text{ is not satisfiable}.
\]
\end{definition}

%
%

It is a trivial observation that a truthful, effective and minimally invasive censor has to answer the same query always in the same way.
\begin{lemma}
Let $\Cens$ be a truthful, effective and minimally invasive censor. Further let $\PK$ be a privacy configuration and $q$ be a query sequence with $q_i = q_j$ for some $i,j$.
Then 
\[
\CensPK(q)_i = \CensPK(q)_j.
\]
\end{lemma}

Consider a truthful, effective, continuous and minimally invasive censor and a given query sequence. If  the censor lies to answer some query, then giving the correct answer would immediately reveal a secret.

\begin{lemma}\label{l:five:1}
Let $\Cens$ be a truthful, effective, continuous and minimally invasive censor. Further let $\PK$ be a privacy configuration and $q$ be a query sequence. Let $i$ be the least natural number such that
\[
\CensPK(q)_i \neq \eval(\KB, q_i).
\]
Let $\CensPK'$ be such that 
\[
\CensPK'(q)|_{i-1} = \CensPK(q)|_{i-1} 
\]
and 
\[
\CensPK'(q)_i = \eval(\KB, q_i).
\]
Then it holds that
\[
\co(\CensPK'(q),i) \models \Box s \quad\text{for some $s \in \Sec$}.
\]
\end{lemma}
\begin{proof}
%
Consider the query sequence $q'$ given by $q'_j := q_j$ for $j<i$ and  $q'_j := q_i$ for $j \geq i$, i.e.~$q'$ has the form
$(q_1,q_2,\ldots,q_{i-1},q_i,q_i,q_i,\ldots)$.
In particular, we have $q|_i = q'|_i$. Thus by continuity of the censor we find
\[
\CensPK(q)|_{i} =  \CensPK(q')|_{i}.
\]

Thus  $\CensPK(q')_i  \neq \eval(\KB, q_i)$. By the definition of minimally invasive we find 
that for some $n$
\begin{equation}\label{eq:min:1}
\co(\CensPK'(q'),n) \models \Box s \quad\text{for some $s \in \Sec$}
\end{equation}
or 
\begin{equation}\label{eq:min:2}
\co(\CensPK'(q'),n) \text{ is not satisfiable}.
\end{equation}
Since the censor is truthful and by Corollary~\ref{cor:cred:1}, we find that \eqref{eq:min:2} is not possible. Thus~\eqref{eq:min:1} holds for some $n$. 

By the definition of $q'$ and the previous lemma we find
\[
\co(\CensPK'(q'),n) = \co(\CensPK'(q'),i) 
\]
if $i \leq n$.
Thus, in case $i \leq n$, \eqref{eq:min:1} implies
\begin{equation}\label{eq:aim:1}
\co(\CensPK'(q),i) \models \Box s \quad\text{for some $s \in \Sec$}.
\end{equation}
In case $i > n$, we find by Lemma~\ref{l:lemmaOne:1} that
\[
\co(\CensPK'(q),n) \subseteq \co(\CensPK'(q),i).
\]
Thus again \eqref{eq:min:1} implies \eqref{eq:aim:1}, which finishes the proof.
\end{proof}

Next we define the notion of a repudiating censor, which garantees that there is always  a knowledge base in which no secret holds and which, given as input to the answering function, produces the same results as the actual knowledge base. Hence this definition provides a version of plausible deniability for all secrets.

\begin{definition}
A censor\/ $\Cens$ is called \emph{repudiating} if{f} for  each privacy configuration $\PK$ and each query sequence $q$, there are knowledge bases $\KB_i$ ($i \in \omega$) such that
\begin{enumerate}
\item $(\KB_i,\AK,\Sec)$ is a privacy configuration for each $i \in \omega$;
\item $\CensPK(q)|_n = \Cens_{(\KB_n,\AK,\Sec)}|_n$, for each $n \in \omega$;
\item $\KB_i \not\vdash_\Lo s$ for each $s\in \Sec$ and each $i \in \omega$.
\end{enumerate}
\end{definition}


Now we can establish our first no-go theorem, which is a generalization of Th.~50 in~\cite{sw14}.

\begin{theorem}[First No-Go Theorem]\label{th:nogo:1}
A continuous and truthful censor satisfies at most two of the properties effectiveness, minimal invasion, and repudiation.
\end{theorem}
\begin{proof}
Let the censor $\Cens$ be continuous, truthful, effective, and minimally invasive. We show that $\Cens$  cannot be repudiating.
We let $S$ be an $\FL$-formula and consider the privacy configuration $\PK$ given by
\[
\KB :=\{S\} \qquad \AK :=\emptyset \qquad \Sec :=\{S\}
\]
and the query sequence $q:=(S,S,\ldots)$.
We set 
\[
a:=\CensPK(q).
\]
Obviously, we have $a = (r,r,\ldots)$ since otherwise $\Cens$ would either
be lying (i.e.~not be truthful) or revealing a secret (i.e.~not be effective).
 
 Now asssume that $\Cens$ is repudiating.
Then there exists a knowledge  base $\KB_1$ such that
\begin{enumerate}
\item $(\KB_1,\AK,\Sec)$ is a privacy configuration;
\item $\CensPK(q)|_1 = \Cens_{(\KB_1,\AK,\Sec)}(q)|_1$;
\item $\KB_1 \not\vdash_\Lo S$.
\end{enumerate}
Let $(a'_1):=\Cens_{(\KB_1,\AK,\Sec)}(q)|_1$. 
Because of  $\KB_1 \not\vdash_\Lo S$ and $\Cens$ being truthful, we find that
$a'_1 = u$ or $a'_1 = r$.  

Suppose towards a contradiction that
\begin{equation}\label{eq:nogo:1}
a'_1 = r.
\end{equation}
Now let $\Cens'$ be a censor as in Lemma~\ref{l:five:1}, i.e.~such that 
\begin{equation}\label{eq:nogo:3}
\Cens'_{(\KB_1,\AK,\Sec)}(q)_1 = u = \eval(\KB_1, S).
\end{equation}
By Lemma~\ref{l:five:1} we get 
\begin{equation}\label{eq:nogo:2}
\co(\Cens'_{(\KB_1,\AK,\Sec)}(q),1) \models \Box S.
\end{equation}
However, by~\eqref{eq:nogo:3} we also have $\co(\Cens'_{(\KB_1,\AK,\Sec)}(q),1) = \{\lnot \Box S\}$, which contradicts~\eqref{eq:nogo:2}.

Hence  \eqref{eq:nogo:1} is not possible and thus we have $a'_1 = u$. This, however, contradicts $\CensPK(q)|_1 = \Cens_{(\KB_1,\AK,\Sec)}(q)|_1$.
We conclude that $\Cens$ cannot be repudiating.
\end{proof}


\section{Non-refusing censors}\label{sec:nonref:1}

In this section we study censors that do not refuse to answer a query. 

\begin{definition}
A censor is \emph{non-refusing} if it never assigns the answer $r$ to a query.
\end{definition}

Of course, a non-refusing censor has to lie in order to keep the secrets. That means if  a censors of this kind shall be effective, then it cannot be truthful. 

Even if we consider lying censors, we work with the assumption that 
\begin{equation}\label{eq:assumption:q}
\text{an attacker believes every answer of the censor.}
\end{equation}
Otherwise, we are in a situation where an attacker cannot believe any answer because the attacker 
does not know which answers are correct and which are wrong, which means that any answer could be a lie.
In that case, querying a knowledge base would not make any sense at all.\footnote{This is, of course, not completely true.  It is possible to distort knowledge bases in such a way that privacy is preserved but statistical inferences are still informative, see, e.g.~\cite{PrivStatInf}.}

Because of the assumption~\eqref{eq:assumption:q}, we can use our notions of effectiveness (Definition~\ref{def:effective:1}) and credibility (Definition~\ref{def:credible:1}) also in the context of lying censors: an attacker should not believe any secret and the beliefs should be satisfiable.
%
%
%

Theorem~\ref{th:nogo:1} about truthful censors did not make any assumptions on the underlying logic~$\Lo$.
The next theorem about non-refusing censors is less general as it is based on classical logic. 
We will use $a,b,c,\ldots$ for atomic propositions and $A, B, C, \ldots$ for arbitrary formulas.

Moreover, we assume that the knowledge base $\KB$ only contains atomic facts (we say $\KB$ is \emph{atomic}). That is if $F \in \KB$, then $F$ is either of the form $p$ or of the form $\lnot p$ where $p$ is an atomic proposition.
Hence we find that if $\KB \vdash_\Lo a \to b$ for two distinct atomic propositions $a$ and $b$, then $\KB \vdash_\Lo \lnot a$ or $\KB \vdash_\Lo b$. We can formalize this using the set of a priori knowledge by letting
\[
\Box (a \to b) \to (\Box \lnot a \lor \Box b) \in \AK.
\] 

Now we can establish our second no-go theorem, which is a generalization of the results of~\cite{Biskup2000}.
\begin{theorem}[Second No-Go Theorem] 
Let $\Lo$ be based on classical logic.
A continuous and non-refusing censor  cannot be at the same time effective and minimally invasive.
\end{theorem}
\begin{proof}
Let the censor $\Cens$ be continuous, non-refusing, and minimally invasive. We show that $\Cens$  cannot be effective.
Let $\Lo$ be classical propositional logic.
We consider the knowledge base 
\[
\KB :=\{a, b\}
\]
where both $a$ and $b$ shall be kept secret, i.e.
\[
\Sec :=\{a, b\}.
\]
Further we assume that it is a priori knowledge that $\KB$ is  atomic. Thus, in particular,
\begin{align*}
\Box (c \to a) \to (\Box \lnot c \lor \Box a) &\in \AK\\
\Box (\lnot c \to b) \to (\Box c \lor \Box b) &\in \AK
\end{align*}
We consider the query sequence $q:=(c \to a, \lnot c \to b , c, \ldots)$ and
set $a:=\CensPK(q)$.

We find 
$\Cens(c \to a) = t$ since $\Cens$ is minimally invasive and $\KB$ might contain $\lnot c$.
Further, we find $\Cens(\lnot c \to b) = t$ since $\Cens$ is minimally invasive and $\KB$ might contain $c$. 

Note that after issuing the first two queries of the sequence~$q$, an attacker knows that $a$ or $b$ must be entailed by~$\KB$. But since the attacker  does not know which one is the case, no secret is leaked.
Formally we have
\begin{align}
\co(\Cens(q),2) &\vdash_\ML  \Box(c \to a)  \label{eq:nogo2:1}
\intertext{and}
\co(\Cens(q),2) &\vdash_\ML  \Box(\lnot c \to b). \label{eq:nogo2:2}
\end{align}
By basic modal logic, \eqref{eq:nogo2:1} and \eqref{eq:nogo2:2} yield
\begin{align}
\co(\Cens(q),2) &\vdash_\ML  \Box c \to \Box a  \label{eq:nogo2:3}
\intertext{and}
\co(\Cens(q),2) &\vdash_\ML  \Box \lnot c \to \Box b,  \label{eq:nogo2:4}
\end{align}
respectively.
Using the a priori knowledge $\AK$, we obtain from \eqref{eq:nogo2:1} and \eqref{eq:nogo2:2}
\begin{align}
\co(\Cens(q),2) &\vdash_\ML  \Box \lnot c \lor \Box a  \label{eq:nogo2:5}
\intertext{and}
\co(\Cens(q),2) &\vdash_\ML  \Box c \lor  \Box b.  \label{eq:nogo2:6}
\end{align}
Because of $\Box c \lor \lnot \Box c$, we get by  \eqref{eq:nogo2:3} and \eqref{eq:nogo2:6} that
\[
\co(\Cens(q),2) \vdash_\ML  \Box a \lor \Box b.
\]
Thus, at this stage, it is known that a secret  holds, but an attacker does not know which one and hence privacy is still preserved.

Now comes the third query, which is $c$. There are two possibilities for a non-refusing censor to choose from:
\begin{enumerate} 
\item $(a)_3 = u$ (which is true). We find $\co(\Cens(q),3) \vdash_\ML \lnot \Box c$. By~\eqref{eq:nogo2:6} we get  
$\co(\Cens(q),3) \vdash_\ML  \Box b$ and a secret is leaked.
\item $(a)_3 = t$ (which is a lie). We find $\co(\Cens(q),3) \vdash_\ML \Box c$. By  \eqref{eq:nogo2:3} we get  
$\co(\Cens(q),3) \vdash_\ML  \Box a$ and a secret is leaked.
\end{enumerate}
In both cases, a secret is leaked. Thus the censor cannot be effective.
%
\end{proof}

To avoid this problem, a censor must not only protect the single elements of $\Sec$ but also their disjunction~\cite{Biskup2000}. For the privacy configuration of the previous proof that means $\Cens$ must also protect $a \lor b$. Then, already the second query, $\lnot c \to b$ would be answered with $u$ because the answer $t$, as shown above, reveals  $a \lor b$. 

Note that protecting the disjunction of all secrets is not as simple as it sounds. Consider, for instance, a hospital information system that should protect the disease a patient is diagnosed with. In this case, protecting the disjunction of all secrets means protecting the information that the patient has some disease. This, however, is not feasible as it is general background knowledge that everybody who is a patient in a hospital has some disease.
Worse than that, sometimes the disjunction of all secrets may even be a logical tautology, which cannot be protected.

\section{Conclusion}

In this paper, we have established two no-go theorems for data privacy using tools from modal logic.
We are confident that logical methods will play  an important role for finding new impossibility theorems or for better understanding already known ones, see, e.g., the logical analyses carried out in~\cite{DBLP:journals/corr/abs-1804-01106} and~\cite{Pacuit2016}.

Another line of future research relates to the fact that  refusing to answer a query can give away the information that there exists a secret that could be infered from some other answer. 
Similar phenomena may occur in multi-agent systems when one of the agents refuses to communicate. 
For example, imagine the situation of an oral exam where the examiner asks a question and the student
keeps silent. In this case the examiner learns that the student does not know the answer to the question for otherwise the student would have answered.

It is also possible that refusing an answer can lead to knowing that someone else knows a certain fact.
Consider the following scenario. A father enters a room where his daughter is playing and he notices that
one of the toys is in pieces. So he asks who has broken the toy. The daughter does not want to betray her
brother (who actually broke it) and she also does not want to lie. Therefore, she refuses to answer her
father's question. Of course, then the father knows that his daughter knows who broke the toy for
otherwise the daughter could have said that she does not know.

We believe that it is worthwhile to study the above situations using general communication protocols that include the possibility of refusing an answer and to investigate the implications of refusing in terms of higher-order knowledge.

\bibliographystyle{abbrv}
\bibliography{mybib}   
\end{document}